\documentclass[twocolumn,showpacs,preprintnumbers,showkeys,superscriptaddress]{revtex4}
\usepackage{amssymb}
\usepackage{graphicx}
\usepackage{dcolumn}
\usepackage{bm}
\usepackage{appendix}

\def\eqref#1{Eq.~(\ref{eq:#1})}

\begin{document}

\title{Validity of the Generalized Density Matrix Method  \\
 for Microscopic Calculation of Collective/Bosonic Hamiltonian}
\author{L. Y. Jia}  \email{jial@nscl.msu.edu}
\affiliation{National Superconducting Cyclotron Laboratory and
Department of Physics and Astronomy, Michigan State University, East
Lansing, Michigan 48824, USA }
\author{V. G. Zelevinsky}
\affiliation{National Superconducting Cyclotron Laboratory and
Department of Physics and Astronomy, Michigan State University, East
Lansing, Michigan 48824, USA }

\date{\today}

\begin{abstract}

Recently a procedure by generalized density matrix (GDM) is proposed
\cite{Jia_2} for calculating a collective/bosonic Hamiltonian
microscopically from the shell-model Hamiltonian. In this work we
examine the validity of the method by comparing the GDM results with
that of the exact shell-model diagonalization in a number of models.
It is shown that the GDM method reproduces the low-lying collective
states quite well, both for energies and transition rates, across
the whole region going from vibrational to $\gamma$-unstable and
deformed nuclei.

\end{abstract}


\vspace{0.4in}

\maketitle

It is a long-standing problem in nuclear physics to understand how
macroscopic collective motion arises from microscopic
single-particle motion. The shell-model (configuration interaction)
successfully reproduces various collective behaviors by
diagonalizing the nucleon Hamiltonian in a huge Slater-determinant
basis. However, the dimension of the basis makes it impractical
except for the cases with only a few valence nucleons. On the other
hand, phenomenological bosonic approaches are often successful in
fitting the experimental data (first of all the geometric Bohr
Hamiltonian \cite{Bohr,Bohr_Hamil} and the interacting boson model
\cite{IBM}). This shows that, out of the huge Slater-determinant
space, there exists a few degrees of freedom with a bosonic nature,
which are usually enough in describing the collective states.
Serious efforts were devoted to deriving those parameters of the
bosonic Hamiltonian from the underlying shell-model Hamiltonian.
However, the complete theory is still missing.

Recently we proposed \cite{Jia_2} a procedure based on the
generalized density matrix (GDM) that was originally formulated in
Refs. \cite{kerman63,BZ,Zele_ptps,shtokman75}. This procedure is
rather simple, clean, and consistent. In compact form, there are
only two equations, (14) and (23) in Ref. \cite{Jia_2}. Results from
the lowest orders give the well-known Hatree-Fock (HF) equations and
random phase approximation (RPA). Higher orders fix the anharmonic
terms in the collective/bosonic Hamiltonian. The aim of this work is
to demonstrate the validity of the GDM method, by comparing its
results with that of the exact shell-model diagonalization.

\begin{figure*}
\includegraphics[height=5in, trim = 0mm 10mm 0mm -100mm, clip, angle=270]{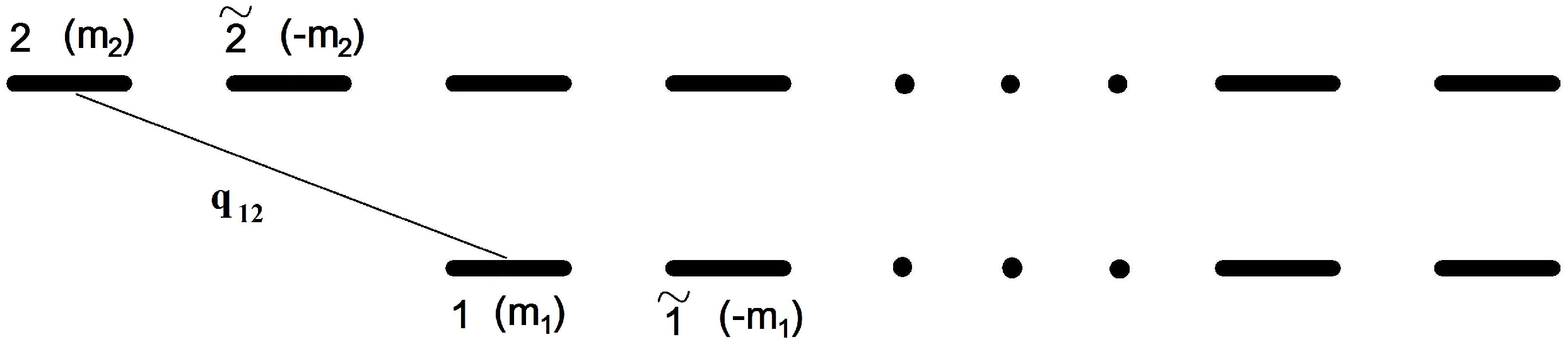}
\caption{\label{Fig_sp} Single-particle level scheme. $\tilde{1}$ is
the time-reversal level of $1$. Each level has a quantum number
$m$.}
\end{figure*}

In this work for simplicity we restrict ourselves to systems without
rotational symmetry. The GDM formulation with angular-momentum
vector coupling has been considered in Ref \cite{Jia_1}. The single
particle (s.p.) space in this work is drawn schematically in Fig.
\ref{Fig_sp}. There are two degenerate s.p. levels with energies $e
= \pm 1/2$. The Fermi surface is in between, thus the lower levels
are completely filled and upper levels are empty. Each s.p. level
has a quantum number $m$ that is a half integer. Degenerate
time-reversal pair has $m$ with different sign, $m_{\tilde{1}} = -
m_1$. For fermions, $|\tilde{\tilde{1}}\rangle = - |1\rangle$, and
we choose the phases such that
\[ |\widetilde{m}\rangle = |-m\rangle ~~~,~~~ |\widetilde{-m}\rangle = -|m\rangle ~~~~~ (m > 0) .   \]
We assume a two-body Hamiltonian,
\begin{eqnarray}
H = \sum_{12} f_{12} a_1^\dagger a_2 + \frac{1}{4} \sum_{1234}
V_{1234} N[a_1^\dagger a_2^\dagger a_3 a_4] ,  \label{H_f}
\end{eqnarray}
where $f_{12} = \delta_{12} e_1$, $e_1$ are the HF s.p. energies
shown in Fig. \ref{Fig_sp}. The density matrix $\rho_{12} =
\delta_{12} n_1$, where the occupation number $n_1 = 1(0)$ for the
lower(upper) s.p. levels. $N[a_1^\dagger a_2^\dagger a_3 a_4]$ is
the normal-ordering form of operators. The residual interaction is
of $Q \cdot Q$ type,
\[ V_{1234} = - \kappa ( q_{14} q_{23} - q_{13} q_{24} ) , \]
where the moment operator $Q = \sum_{12} q_{12} a_1^\dagger a_2$ is
Hermitian and time-even. For simplicity we assume $q$ is real, thus
\[ q_{12} = q_{21} = q_{\tilde{2}\tilde{1}} = q_{\tilde{1}\tilde{2}} . \]
Operator $q$ has certain selection rule with respect to quantum
number $m$, which will be specified later. We further set diagonal
matrix elements of $q$ to zero, $q_{11} = 0$; hence the moment of
the mean field vanishes, $Q^{(00)} = {\rm{Tr}}[q \rho] = 0$.

Following the procedure in Ref. \cite{Jia_2}, we are able to map the
fermionic Hamiltonian (\ref{H_f}) onto a bosonic Hamiltonian
\begin{eqnarray}
H_b = \omega^2 \frac{\alpha^2}{2} + \frac{\pi^2}{2} + \Lambda^{(30)}
\frac{\alpha^3}{3!} + \Lambda^{(12)} \frac{\{\alpha, \pi^2\}}{4}
+ \Lambda^{(40)} \frac{\alpha^4}{4!}  \nonumber \\
+ \Lambda^{(22)} \frac{\{ \alpha^2 , \pi^2 \}}{8}  + \Lambda^{(04)}
\frac{\pi^4}{4!} + \Lambda^{(50)} \frac{\alpha^5}{5!} + \ldots , ~~~
\label{H_b}
\end{eqnarray}
where the collective coordinate $\alpha$ and momentum $\pi$ satisfy
$[\alpha,\pi] = i$. Meanwhile, the generalized density matrix
$r^{(mn)}$ in the expansion
\begin{eqnarray}
R_{12} \equiv a_2^\dagger a_1 = \rho_{12} + r^{(10)}_{12} \alpha +
r^{(01)}_{12} \pi + r^{(20)}_{12} \frac{\alpha^2}{2} + \ldots
\label{R_exp}
\end{eqnarray}
is expressed in terms of $\Lambda^{(mn)}$. The bosonic Hamiltonian
$H_b$ (\ref{H_b}) should reproduce the low-lying collective spectrum
of the original nucleon Hamiltonian $H$ (\ref{H_f}). Substituting
the solution (\ref{R_exp}) into $Q = \sum_{12} q_{12} a_1^\dagger
a_2$, we get the boson image of the latter,
\begin{eqnarray}
Q_b = Q^{(10)} \alpha + Q^{(20)} \frac{\alpha^2}{2} + Q^{(02)}
\frac{\pi^2}{2} + Q^{(30)} \frac{\alpha^3}{3!} + \ldots , ~~~
\label{Q_b}
\end{eqnarray}
where $Q^{(mn)} = {\rm{Tr}}[q r^{(mn)}]$, and time-odd terms vanish
automatically. The transition rates calculated from $Q_b$ between
eigenstates of $H_b$ should reproduce that of $Q$ between
eigenstates of $H$.

As stated in Ref. \cite{Jia_2}, the GDM method fixes $H_b$
completely. In each even order (quadratic, quartic ...) in $H_b$,
the GDM method gives one constraint on $\Lambda^{(mn)}$'s. The
number of constraints is the same as that of independent parameters
in $H_b$, removing in Eq. (\ref{H_b}) superficial degrees of freedom
due to canonical transformations of $\alpha$ and $\pi$ conserving
$[\alpha,\pi] = i$.\\

In the following we consider four models with different structures
(different configurations of s.p. levels and different selection
rules of $q$). We start with the simplest case of the first model.
Both the upper and lower group have $12$ degenerate s.p. levels with
quantum numbers $m = \pm \frac{1}{2}, \pm \frac{3}{2}, ... , \pm
\frac{11}{2}$. Operator $q$ has the selection rule $\Delta m = 0$,
that is, $q_{12}$ vanishes unless $m_1 = m_2$. The non-vanishing
$q_{12}$ ($m_1 = m_2$) are set to be $1$.

In this model we find by numerical computation a ``symmetry''. That
is, there are only three non-vanishing terms in Hamiltonian
(\ref{H_b}): $\omega^2$, $\Lambda^{(40)}$, and $\Lambda^{(22)}$
(besides $\pi^2/2$). We suspect there was some kind of ``quasi
angular-momentum symmetry'', similar to that in the Lipkin model,
where the only three non-vanishing terms are $\omega^2$,
$\Lambda^{(40)}$, and $\Lambda^{(04)}$ (see Ref. \cite{Jia_2}).

\begin{figure*}
\includegraphics[width = 1.0\textwidth]{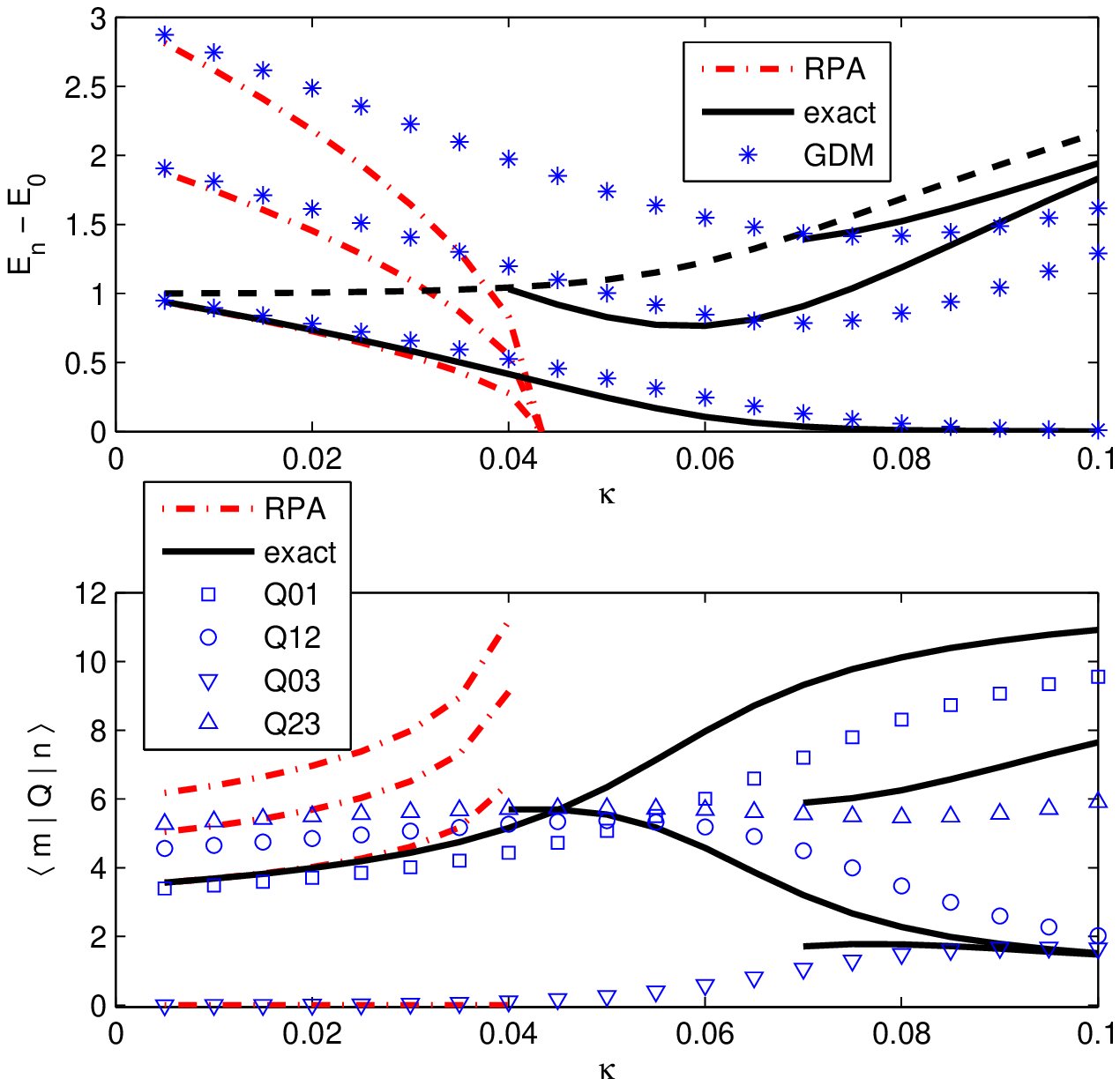}
\caption{\label{Fig_delta0} (Color online) Excitation energies $E_n
- E_0$ and transition matrix elements $\langle m | Q | n \rangle$ in
the first model of this work as a function of $\kappa$. The black
solid lines are exact results by shell-model diagonalization. The
black dashed line is the beginning of the single particle continuum.
The red dashed-dotted lines are the RPA results. The blue symbols
are GDM results. The asterisks are energies; and the squares,
circles, up-triangles and down-triangles are matrix elements of $Q$
between different states. $\langle m | Q | n \rangle$ that are not
shown vanish in both the shell model and the GDM calculations.}
\end{figure*}

The results are shown in Fig. \ref{Fig_delta0}. We see that the GDM
calculation reproduces the exact results of the shell model quite
well, both for energies and transition rates. In the shell model we
calculate the lowest several states by Lanczos method. The dashed
line in the upper panel is the beginning of the s.p. continuum, only
those collective states below the continuum can be seen. In the GDM
calculation the resultant bosonic Hamiltonian is diagonalized in a
finite ``physical'' bosonic space, $\{| 0 \le n \le 12 \rangle\}$.
$|n\rangle$ is the $n$ phonon state, $A^\dagger A |n\rangle = n
|n\rangle$, $A^\dagger = ( u \alpha + i v \pi )/\sqrt{2} , u v = -
1$. The coefficient $u$ is fixed by minimizing $A |\rm{HF}\rangle$
in its one-particle-one-hole components, where $|\rm{HF}\rangle$ is
the Hartree-Fock ground state that is mapped onto the bosonic state
$|n=0\rangle$. The result is $u^4 = \frac{ \sum_{2<F<1}
|r^{(10)}_{12}|^2 }{ \sum_{2<F<1} |r^{(01)}_{12}|^2 }$, where the
summation indices $1$ and $2$ run over unoccupied and occupied s.p.
levels, respectively (``$F$'' means Fermi surface). In models of
this work, $u$ is a number close to $1$. The shown GDM energies and
transitions are practically independent of small variations of $u$
around the above value.

As $\kappa$ increases, the system goes from vibrational to
$\gamma$-unstable region. In the vibrational region, higher excited
states are influenced more by the anharmonicities, as expected. At
large $\kappa$ the spectrum becomes doubly degenerate in a deep
double-well potential (large negative $\omega^2$).

\begin{figure*}
\includegraphics[width = 1.0\textwidth]{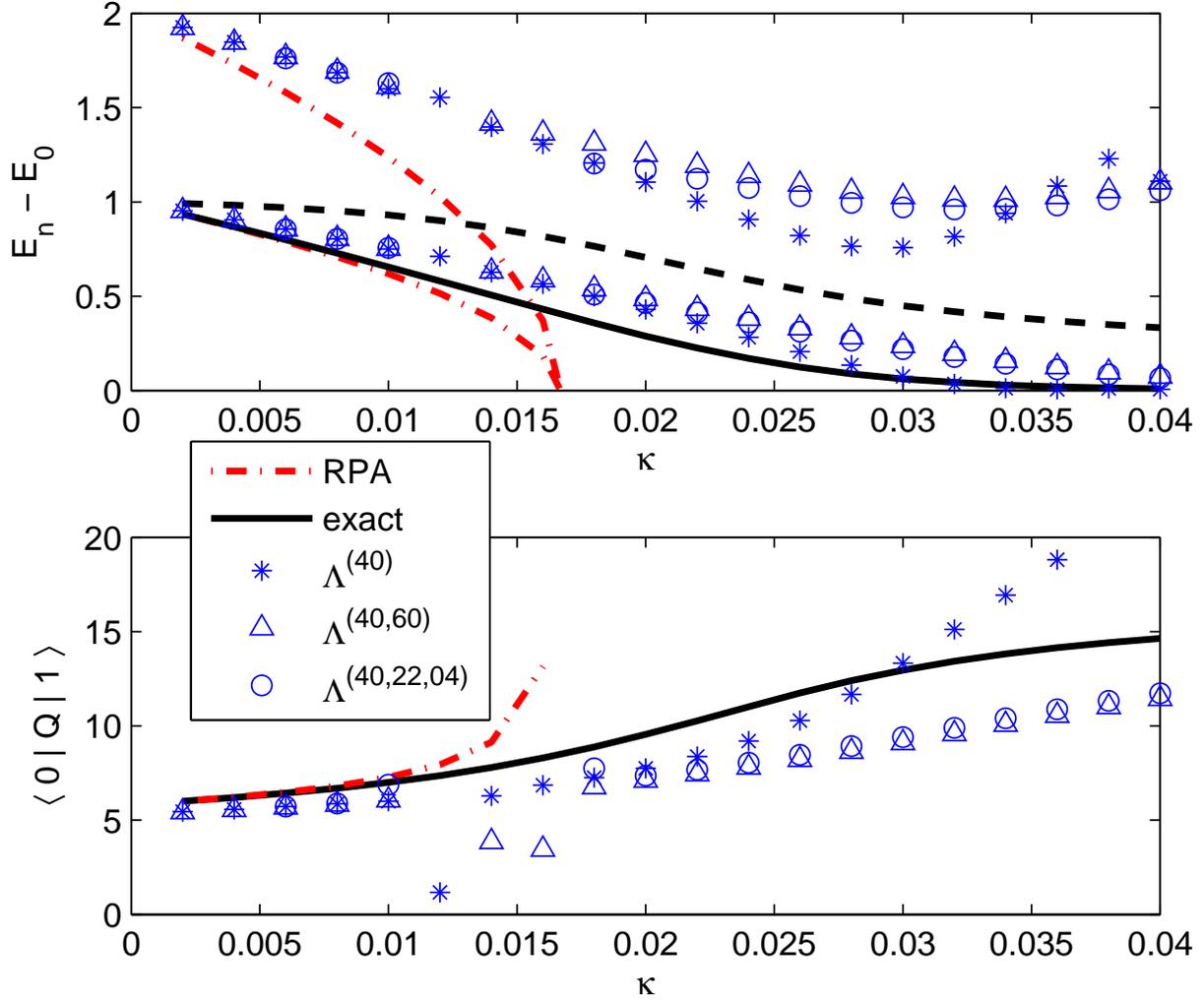}
\caption{\label{Fig_delta01} (Color online) Excitation energies $E_n
- E_0$ and transition matrix elements $\langle 0 | Q | 1 \rangle$ in
the second model of this work as a function of $\kappa$. The black
solid lines are exact results by shell-model diagonalization. The
black dashed line is the beginning of the single particle continuum.
The red dashed-dotted lines are the RPA results. The blue symbols
(asterisks, up-triangles and circles) are results of three different
sets of GDM calculations, as labeled in the legend. }
\end{figure*}

\begin{figure*}
\includegraphics[width = 1.0\textwidth]{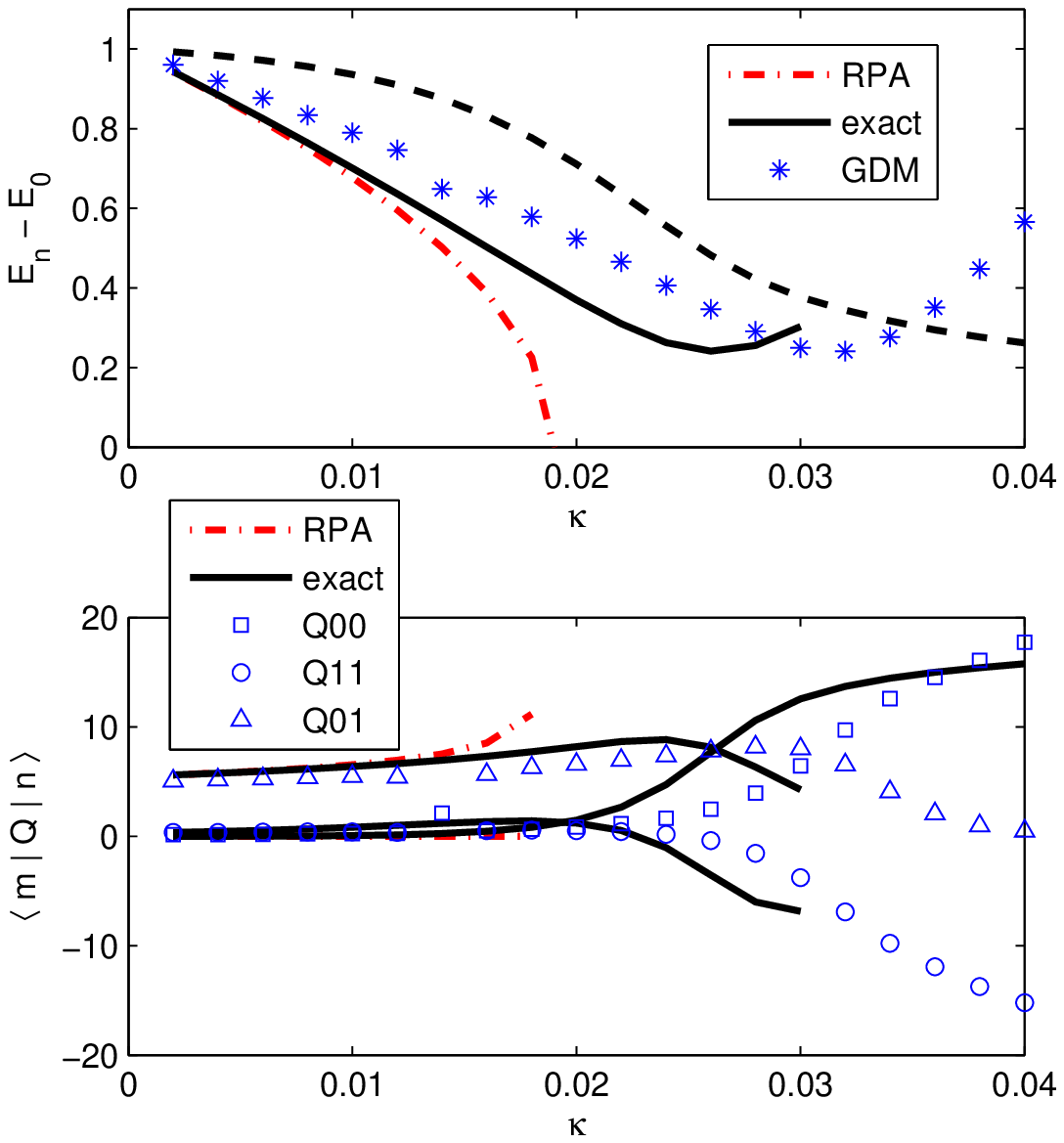}
\caption{\label{Fig_prolate} (Color online) Excitation energies $E_n
- E_0$ and transition matrix elements $\langle m | Q | n \rangle$ in
the third model of this work as a function of $\kappa$. The black
solid lines are exact results by shell-model diagonalization. The
black dashed line is the beginning of the single particle continuum.
The red dashed-dotted lines are the RPA results. The blue symbols
are GDM results. The asterisks are energies; and the squares,
circles and up-triangles are matrix elements of $Q$ between
different states.  }
\end{figure*}

\begin{figure*}
\includegraphics[width = 1.0\textwidth]{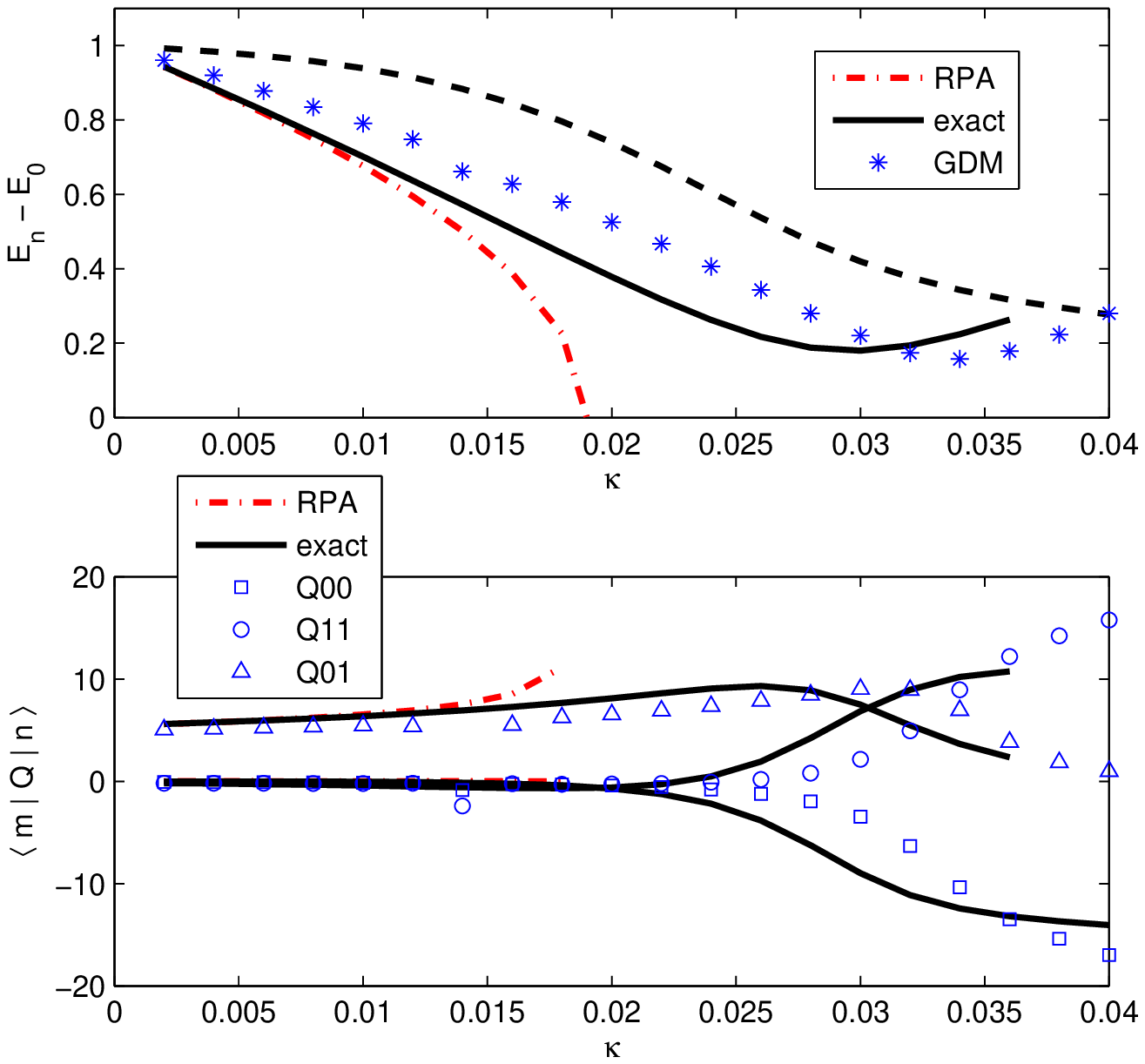}
\caption{\label{Fig_oblate} (Color online) Excitation energies $E_n
- E_0$ and transition matrix elements $\langle m | Q | n \rangle$ in
the fourth model of this work as a function of $\kappa$. The black
solid lines are exact results by shell-model diagonalization. The
black dashed line is the beginning of the single particle continuum.
The red dashed-dotted lines are the RPA results. The blue symbols
are GDM results. The asterisks are energies; and the squares,
circles and up-triangles are matrix elements of $Q$ between
different states. }
\end{figure*}

We would like to mention an important point, that the GDM method
works better with increasing $\Omega$ (collectivity). Another
calculation has been done (but not shown here) with $8$ particles in
$16$ s.p. levels. The GDM results of the current calculation ($12$
particles in $24$ s.p. levels) have very clear improvement over that
of the former. In other words, the error in Fig. \ref{Fig_delta0}
may be of order $1/\Omega$. The largest part of this error may come
from the RPA frequency $\omega^2$. At the current stage, the GDM
method calculates all $\Lambda^{(mn)}$ in their leading order of
$1/\Omega$ but not the next. We suspect that the correct $\omega^2 =
\Lambda^{(20)}$ is smaller by a $1/\Omega$ term than the one
determined here by the RPA equation. This would shift all the GDM
curves to the left (smaller $\kappa$) a little bit, which would
decrease greatly the systematic error (see Fig. \ref{Fig_delta0}).
This systematic error own to $\omega^2$ seems to be present in all
models in this work (see Figs. \ref{Fig_delta01}-\ref{Fig_oblate}).
Also, it is confirmed in the Lipkin model where everything is known
analytically (see Ref. \cite{Jia_2}). Hence an apparent improvement
would be calculating $\omega^2$ in its next-to-leading order of
$1/\Omega$.

Next we consider the second model, which has the same s.p.
configuration but a different $q$ that now has the selection rule
$\Delta m = 0 , \pm 1$. Non-vanishing $q_{12}$ is still set to be
$1$. In this model we did not find a ``symmetry'' as in the previous
case, so the problem exists of what should be the ``best'' mapping.
In the following we did three sets of GDM calculations. The first
calculation keeps only $\Lambda^{(40)}$ (besides $\omega^2
\alpha^2/2$ and $\pi^2/2$) in $H_b$, which is fixed by the
constraint from $4$th order in e.o.m. The second calculation keeps
only $\Lambda^{(40)}$ and $\Lambda^{(60)}$, which are fixed by the
two constraints from upto $6$th order in e.o.m. The third
calculation keeps only $\Lambda^{(40)}$, $\Lambda^{(22)}$, and
$\Lambda^{(04)}$, fixed by the three constraints from upto $8$th
order in e.o.m.

We first notice in Fig. \ref{Fig_delta01} that in this model the
s.p. continuum goes down as increasing $\kappa$, as opposed to going
up in the previous model. This is because now mixing of s.p. levels
within the upper(lower) group is allowed by the selection rule that
$\Delta m$ can be $\pm 1$. As a result, originally degenerate levels
from the upper(lower) group get a finite spread, which decreases the
gap of the s.p. continuum. Only the first excited state is below the
gap and calculated in the shell model.

In the GDM calculations we see that the simplest one
degree-of-freedom ($\Lambda^{(40)}$) calculation is reasonably well
in most cases except at very large $\kappa$. The other two
calculations ($\Lambda^{(40/60)}$ and $\Lambda^{(40/22/04)}$) give
essentially the same results (for the quantities shown in Fig.
\ref{Fig_delta01}), although their common parameter,
$\Lambda^{(40)}$, differ a lot. This insensitivity of GDM results to
the degrees of freedom chosen, is important. As we said, two
different bosonic Hamiltonians could be equivalent if they were
related by canonical transformations/renormalizations of variables
$\alpha$ and $\pi$. This insensitivity simply says that the GDM
formulism knows these renormalizations and does them correctly. In
the first model we also find this insensitivity (but not shown).
Finally we notice that in regions of $\omega^2 \sim 1/\Omega$,
calculations that go to higher orders in e.o.m. may give unphysical
results. This is again because the e.o.m. are accurate in leading
order of $1/\Omega$ but not the next. The fact that this
``divergence'' appears slightly before the instability point of RPA
shown in Fig. \ref{Fig_delta01}, indicates again that the correct
$\omega^2$ may be smaller than the one calculated by RPA, as
mentioned before.

At last we consider two models with s.p. configuration that is
asymmetric in upper and lower groups, which generates odd
anharmonicities that are necessary for deformation. In the third
model, the lower group has $10$ s.p. levels with $m = \pm
\frac{3}{2}, ... , \pm \frac{11}{2}$, the upper group has $14$ s.p.
levels with $m = \pm \frac{1}{2}, ... , \pm \frac{13}{2}$. In the
fourth model, the lower group has $12$ s.p. levels with $m = \pm
\frac{1}{2}, ... , \pm \frac{11}{2}$, the upper group has $10$ s.p.
levels with $m = \pm \frac{1}{2}, ... , \pm \frac{9}{2}$. In both
models, operator $q$ still has the selection rule of $\Delta m = 0 ,
\pm 1$, with non-vanishing matrix elements set to be $1$. The third
model has a slightly larger asymmetry than that of the fourth model,
and their signs of the asymmetry are different.

These two models are more complicated in the sense that now there
are more active degrees of freedom (odd anharmonicities). In the GDM
method, we do a possibly simplest calculation. We keep in $H_b$ only
$\Lambda^{(30)}$, $\Lambda^{(12)}$, and $\Lambda^{(40)}$ (besides
$\omega^2 \alpha^2/2$ and $\pi^2/2$). $\Lambda^{(30)}$ and
$\Lambda^{(12)}$ are fixed by requiring $Q^{(20)} = Q^{(02)} = 0$ in
the solution (\ref{Q_b}). Then $\Lambda^{(40)}$ is fixed by the
constraint from $4$th order in e.o.m. The requirement $Q^{(20)} =
Q^{(02)} = 0$ is the same as that for previous two models without
upper-lower asymmetry, by which $\Lambda^{(30)}$ and
$\Lambda^{(12)}$ vanish.

The results are shown in Figs. \ref{Fig_prolate}-\ref{Fig_oblate}.
The deformation begins around the critical point of RPA when
$\omega^2$ becomes negative. In the vibrational region the potential
is stiff and deformation is not easy. As $\kappa$ increases, the
potential becomes flat in bottom and finally of a double-well shape.
Then, even a relatively small odd anharmonicity (here mainly
$\Lambda^{(30)}$) can tilt the potential and generate large
deformation. We notice firstly that the GDM calculations give the
correct sign of deformations: positive/negative for the
ground/first-excited state of the third model, and vice versa for
the fourth model. In realistic situation $\Lambda^{(30)}
((\hat{\alpha} \times \hat{\alpha})^2 \times \hat{\alpha})^0 \sim
\Lambda^{(30)} \beta^3 \cos 3\gamma$ ($\hat{\alpha}$ is the
quadrupole phonon and $\beta$, $\gamma$ are Bohr angles), the sign
of $\Lambda^{(30)}$ ``determines'' the intrinsic shape of the
nucleus (prolate or oblate). This is especially interesting in the
transitional regions where the rotor formula is not applicable.
Secondly, the quantitative agreement of deformation is also good
except at the largest $\kappa$. There the deformation ``saturates''
towards its maximal possible value within the model space, favored
by energy. Meanwhile in the boson mapping, we are too close to the
boundary of the finite physical bosonic space, and the GDM results
become inaccurate. In realistic nuclei this ``saturation'' may not
be usual. The number of participating/active nucleons is usually
around $30$ in well-deformed nuclei, which is much larger than that
around $10$ in the current models. Finally, we would like to point
out that the first excited state in our simple models is not a
``rotational'' state, rather it corresponds to the next ``band
head'' in realistic rotational nuclei. The rotational states that
are very low in energy
come in only in three dimensions.\\

In this work we demonstrate the validity of the GDM procedure in
microscopically calculating the collective/bosonic Hamiltonian. The
lowest several states of this bosonic Hamiltonian quite well
reproduce the collective states of the exact shell-model, both for
energies and transition rates, in a wide range from vibrational,
$\gamma$-unstable, to deformed nuclei. Specifically, we show that
deformation can be described without introducing a deformed mean
field. The traditional procedure of ``symmetry breaking and
restoration'', first ``statically'' breaks rotational symmetry in
the ground state, by representing the latter as a Slater determinant
of deformed s.p. levels (Nilsson levels); then projects afterwards
to good angular momentum. However, in case of large shape
fluctuations (flat minimal of energy surface) or shape coexistence
(two close minimal), it may fail. On the other hand, the GDM
procedure always conserves the rotational symmetry. Deformations are
put in ``dynamically'' at higher orders (for example cubic terms)
beyond the mean field. Thus it is suitable to describe for example
shape fluctuation and coexistence problems.

In realistic nuclei, the gap of s.p. continuum is generated by the
pairing correlations. The GDM formulism based on the
Hartree-Fock-Bogoliubov transformation is straightforward and has
been done in Ref. \cite{Jia_1}. However, another treatment may be
possible. Instead of introducing Bogoliubov quasi-particles and
representing the ground state as a Slater determinant of the former,
the pairing correlations are considered in higher orders beyond the
mean field, by keeping both the particle-hole and particle-particle
channels in the factorization $a_4^\dagger a_3^\dagger a_2 a_1
\approx a_4^\dagger a_1 \cdot a_3^\dagger a_2 - a_4^\dagger a_2
\cdot a_3^\dagger a_1 + a_4^\dagger a_3^\dagger \cdot a_2 a_1$. In
this way the symmetry of exact particle number is always conserved.
Work along this line is in progress and results seem promising. We
are also generalizing the GDM code by including angular-momentum
vector coupling that are necessary for realistic calculations.

This work is supported by the NSF grants PHY-0758099 and
PHY-1068217.

\newpage

~

\newpage
\phantom{a}
\newpage

~

\newpage
\phantom{a}
\newpage

~

\newpage
\phantom{a}
\newpage

\newpage

~

\end{document}